\documentclass[conference]{IEEEtran}
\IEEEoverridecommandlockouts
\usepackage{cite}
\usepackage{amsmath,amssymb,amsfonts}
\usepackage{algorithmic}
\usepackage{graphicx}
\usepackage{textcomp}
\usepackage{caption,subcaption}
\usepackage{bm}
\usepackage{tikz}
\usepackage[all]{xy}
\usetikzlibrary{arrows,shapes}
\usetikzlibrary{decorations.pathmorphing}
\usetikzlibrary{decorations.markings}
\usepackage{url}
\usepackage{xcolor}

\def\BibTeX{{\rm B\kern-.05em{\sc i\kern-.025em b}\kern-.08em
    T\kern-.1667em\lower.7ex\hbox{E}\kern-.125emX}}
\begin{document}

\title{The geometry of syntax and semantics for directed file transformations\\
\thanks{The authors thank Sergey Bratus, Arquimedes Canedo, Dan Kaminsky, Bartosz Milewski, David Spivak, and Matvey Yutin for helpful comments. This material is based upon work partially supported by the Defense Advanced Research Projects Agency (DARPA) SafeDocs program under contract HR001119C0072. Any opinions, findings and conclusions or recommendations expressed in this material are those of the author(s) and do not necessarily reflect the views of DARPA.
}
}

\author{\IEEEauthorblockN{Steve Huntsman}
\IEEEauthorblockA{\textit{BAE Systems FAST Labs}\\
Arlington, Virginia, USA \\
steve.huntsman@baesystems.com}
\and
\IEEEauthorblockN{Michael Robinson}
\IEEEauthorblockA{\textit{American University}\\
Washington, DC, USA \\
michaelr@american.edu}
}

\maketitle

\begin{abstract}
We introduce a conceptual framework that associates syntax and semantics with vertical and horizontal directions in principal bundles and related constructions. This notion of geometry corresponds to a mechanism for performing goal-directed file transformations such as ``eliminate unsafe syntax'' and suggests various engineering practices.
\end{abstract}


\section{\label{sec:Introduction}Introduction}

There is a long tradition of considering syntax and semantics as dual, e.g., with various type theories and sorts of categories respectively inhabiting these roles \cite{nlab:relation_between_type_theory_and_category_theory}. Meanwhile, there is an even longer tradition of considering algebra and geometry as (Isbell, ``sheafily,'' or ``spectrally'') dual, as manifested in the notions of Stone duality between Boolean algebras and Stone spaces; or Gel'fand duality between commutative C*-algebras and locally compact Hausdorff spaces; or duality between commutative rings and affine schemes \cite{nlab:isbell_duality}. 
\footnote{
Of particular interest for our present considerations is the duality between crossed product C*-algebras and principal bundles \cite{phillips1984crossed}, and more generally dualities that embody noncommutative geometry \cite{connes1994noncommutative}.
}
However, it can be argued that ``the duality between syntax and semantics is really a manifestation of that between algebra and geometry'' \cite{awodey2013first}.

Here, we introduce a conceptual framework that simultaneously embraces and extends this perspective: syntax transformations are viewed as forming a group (or more generally, a groupoid), and semantically distinct (representations of) files form a ``base space.'' 
\footnote{
The \emph{homotopy hypothesis/theorem} that ``spaces are $\infty$-groupoids'' \cite{univalent2013homotopy} can and should be viewed in this context as giving a geometrical interpretation for this base space, rather than an algebraic interpretation for paths on this base space.
}
These constructs are unified in the structure of a bundle (or more generally, a fibration), and a goal-directed transformation of files imbues this structure with a notion of geometry that connects syntax and semantics.

Perhaps the most contentious part of this conceptual framework is the notion that syntax transformations are (or ought to be) invertible. On one hand, the requirement of vertical invertibility is \emph{imposed} by mathematics once one commits to the model of a principal bundle or, more generally, a category fibered in groupoids (see \S \ref{sec:PrincipalBundles} or \S \ref{sec:ModuliSpaces}, respectively) for providing an arena where geometry can direct file transformations (or conversely, where directed file transformations can be considered as defining a geometry). 
\footnote{
The models we consider might admit further generalization to circumvent this invertibility requirement via monoidal fibrations \cite{moeller2018monoidal}. However, the relative simplicity of the considerations of \S \ref{sec:Bundles} could (and probably should) be taken as a hint to take the invertibility requirement seriously. Indeed, exposing the constraints imposed by a mathematical model is the primary benefit of an exercise such as ours. 
}
On the other hand, the requirement is \emph{justified} by performing transformations with ancillary memory: i.e., annotating any transformations with inline comments or external ancillae that are invertible (into nothingness) by (de)construction. 
\footnote{
This is essentially the reverse of the tactic underlying \emph{logically reversible computation}, for which see, e.g. \cite{bennett1973logical,toffoli1908reversible,kaarsgaard2017logic,morita2017theory}. In reversible computation, one accepts ancillary inputs that are transformed into ``garbage'': here, we accept ancillary outputs that can be reverse-transformed into nothingness. 
}

For example, consider a PDF file \cite{ISO32000}. As \S 2 of \cite{whitington2011pdf} points out, the program \texttt{pdftk} will produce a valid PDF from a malformed file with an abbreviated header and missing data about both the length of the page content stream and the cross-reference table. At the same time, the program will manipulate abstractly irrelevant details of concrete syntax such as whitespace. Insofar as this process would be instantiated in our framework, and notwithstanding the fact that the original malformed and valid PDF files are presented in the same syntactic representation, the overall transformation itself should be considered as the composition of one explicitly invertible ``vertical''/``syntactic'' and one not explicitly invertible ``horizontal''/``semantic'' transformation. The former transformation merely inserts comments detailing the malformations (including tags that detail the comments' provenance and hence facilitate their removal or ``inversion''), whereas the latter transformation actually manipulates the header, inserts the missing data, and performs some collateral manipulation of concrete syntax.

A simpler and completely explicit example in the same vein would be a PDF file with invalid terminal object delimiters, e.g. \texttt{objend} instead of \texttt{endobj}. 
\footnote{
We thank P. Wyatt for noting that this delimiter is invalid, though it can be seen on occasion ``in the wild.''
}
Here, the transformation sequence would be something like
\begin{align}
\texttt{objend} & \Rightarrow^{vert/}_{synt} & \texttt{objend \% objend -> endobj} \nonumber \\
& \Rightarrow^{horz/}_{sema} & \texttt{endobj \% objend -> endobj} \nonumber
\end{align}
which actually specifies how to perform the inverse syntactic transformation.  (Recall that
comments in PDFs are initialized by \texttt{\%} and terminated by end-of-line markers outside of strings or inside of content streams (see \S 7.2.4 of \cite{ISO32000}). 

For lack of a better term, we call this sort of bookkeeping \emph{sugar-neutral}: i.e., we view syntactic sugar such as variable forms for a token in a file or a \texttt{case} or \texttt{elseif} statement in source code as something that should be accomodated and potentially preserved in a file at the \emph{outset} of a transformation, but that should not be introduced \emph{during} a transformation. 
Taking (de)compilation as an example and insisting on \emph{normal forms} as in Figures \ref{fig:loops} and \ref{fig:NoBranches}, we can in principle carry along the concrete syntax of a file through quite complex transformations to detail how to transform between semantically equivalent files with different concrete syntax in any representation (see \S \ref{sec:FileTransformationImplementations}).

\begin{figure}
\centering
\begin{subfigure}[b]{.45\columnwidth}
\begin{small}
\begin{verbatim}
int i;
for (i=0; i<10; i++)
{
  z+=i;
}
\end{verbatim}
\end{small}
\caption{A \texttt{for} loop.}
\end{subfigure} \quad 
\begin{subfigure}[b]{.45\columnwidth}
\begin{small}
\begin{verbatim}
int n=0;
while (n<10) {
  x+=n;
  n++;
}
\end{verbatim}
\end{small}
\caption{A \texttt{while} loop.}
\end{subfigure}
\caption{\label{fig:loops}
(Adapted from \cite{lacomis2019dire}.) Two semantically identical C loops. The more versatile \texttt{while} construct is better suited for a normal form of C (and for a bootstrapping compiler).
}
\end{figure}

\begin{figure}
\centering
\begin{subfigure}[b]{.45\columnwidth}
\begin{small}
\begin{verbatim}
      jmp @5
@4:
      jmp @9
@8:
      jne @19
      jmp @10
@19:
      jmp @14
@13:@14:
      jg  @13
@9:@10:
      jge @20
      jmp @8
@5:@20:
      jge @21
      jmp @4
@21:
\end{verbatim}
\end{small}
\caption{Branches and labels from assembler listing of Gauss-Jordan elimination code.}
\end{subfigure} \quad 
\begin{subfigure}[b]{.45\columnwidth}
\begin{small}
\begin{verbatim}
START; S
do while b
  S
  do while b
    if b
      S
      do while b
        S
      enddo
    endif
    S
  enddo
  S
enddo; HALT
\end{verbatim}
\end{small}
\caption{Algorithmically restructuring control flow from (a) to turn backward branches into \texttt{while} loops. Each \texttt{S} is its own statement/subroutine; each \texttt{b} is its own Boolean predicate.}
\end{subfigure}
\caption{\label{fig:NoBranches}
(Adapted from \cite{zhang2004using}.) Decompiling into normal form.
}
\end{figure}

\section{\label{sec:Bundles}Bundles}

\subsection{\label{sec:PrincipalBundles}Principal bundles}

\begin{figure}[htbp]
\begin{center}
\includegraphics[trim = 45mm 105mm 45mm 90mm, clip, width=\columnwidth,keepaspectratio]{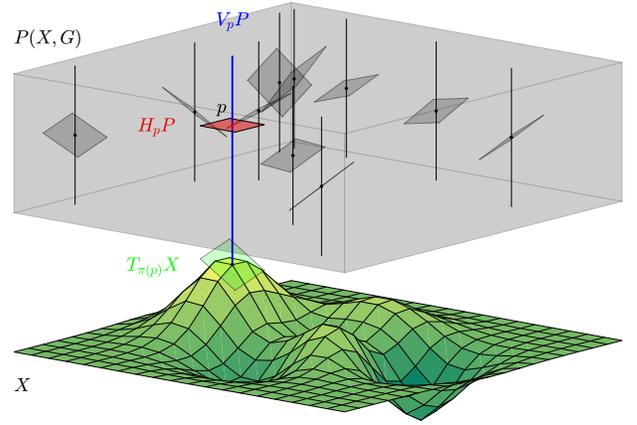}
\end{center}
\caption{ \label{fig:PrincipalBundle} 
A principal $G$-bundle $P(X,G)$ provides a natural arena for geometry realized through a \emph{connection}, i.e., a smooth direct sum decomposition $T_p P = {\color{blue}V_p P} \oplus {\color{red}H_p P}$ of tangent spaces into {\color{blue}``vertical''} and {\color{red}``horizontal''} components that is \emph{equivariant} under an action of $G$. In the figure, $\pi : P \rightarrow X$ is the bundle projection map. 
} 
\end{figure}

The geometry of \emph{principal bundles} \cite{kobayashi1963foundations, lam2015topics} turns out to provide a useful conceptual framework for reasoning about and manipulating syntax and semantics. Given a ``horizontal base space'' $X$ corresponding to some particular lossless representation of a language/file format (e.g., strings/words, concrete syntax trees [CSTs], etc.) and a ``vertical'' group $G$ of invertible syntactic transformations, we consider an object akin to a \emph{connection} in a \emph{principal bundle} $P(X,G)$ as depicted in Figure \ref{fig:PrincipalBundle}. 

In the present context, we can think of a connection as a recipe for directing file transformations in terms of vertical syntactic transformations and horizontal semantic transformations, or in a complementary sense to connect the spaces of lifts of nearby files. Since syntactic file transformations can be thought of as abstractions of semantic file transformations \cite{cousot2002systematic},
the connection (in particular, its equivariance property) informs the distinction between syntax and semantics.
 
In other words, the common notion of lifting a file to a different representation is consistent with the usage of the term ``lift'' in mathematical use: a directed semantic modification corresponds to the terminal point of a (presumably fairly ``short'') path on the base space $X$ that can be lifted to a path in the bundle $P(X,G)$ using a connection. In this differential-geometrical analogy, the goal of semantic file transformations is manifested by side information (e.g., format specifications, corpora, programs that nominally accept such files as inputs, etc.) that define a geometry, and a directed file transformation itself corresponds to \emph{parallel transport} along a vector field. 
\footnote{
It bears mentioning that the relevant mathematics can be adapted to the cases where $X$ and/or $G$ are discrete: these are respectively exemplified by lattice gauge theory \cite{rothe2005lattice} 
and/or discrete gauge groups in physics \cite{dimakis1994differential}. 
The essential idea is obvious: such discrete spaces are embedded in continuous ones that they approximate.
}

An explicit objective function that gives rise to the geometry can be specified in practice by, e.g., a dissimilarity measure between the original and perturbed files \cite{schulte2018evolving}, 
or between traces or indeed any other suitable artifacts (see \S \ref{sec:FibrationGeometry}). In this analogy, ``infinitesimal'' transformations are composed, and the transformation approach of \S \ref{sec:FileTransformations} draws from this conceptual framework.

Crucially, adding or removing sugar in syntactic transformations affects the group structure required to define a notional connection (though as we discuss later, groupoids can suffice for algebraic purposes and a connection \emph{per se} is unnecessary since geometry can be constructed via alternative means). We avoid this by choosing representative semantics-preserving transformations based on \emph{normal forms}. This reflects the observation of \cite{pombrio2014resugaring}
that ``sugaring is a compilation process.'' For programming languages, such normal forms can be enforced by, e.g., restructuring code so that backward branches become \texttt{while} loops \cite{zhang2004using}
or in reference to a particular configuration of some machine learning algorithm such as \cite{lacomis2019dire} (i.e., we fix a trained implementation once and for all).

Under this analogy, in the file transformation process, points in $P(X,G)$ correspond to CSTs and $G$ corresponds to semantics-preserving (invertible) transformations on CSTs. The equivalence class of CSTs that correspond to a given abstract syntax tree (AST) carries both group-theoretical and language-security theoretical significance, as it indicates redundancy in a format.

\subsection{\label{sec:FileTransformations}File transformations}

In order to indicate the substantive nature of the analogy in \S \ref{sec:PrincipalBundles}, we proceed to sketch some examples of directed file transformations in \S \ref{sec:ExampleFileTransformations}, detail the first of these in \S \ref{sec:DetailedExample}, and in \S \ref{sec:FileTransformationImplementations} outline a generic framework for implementing directed file transformations that separates concerns between syntax and semantics.

\subsubsection{\label{sec:ExampleFileTransformations}Examples of directed file transformations}
\footnote{
There may be relatively simple and useful examples obtainable by, e.g., manipulating casts/encodings of simple datatypes.
}
\begin{itemize}
\item Toy example 1 (detailed and elaborated upon in \S \ref{sec:DetailedExample}):
	\begin{itemize}
	\item $X = \{0, 32,\dots,126\}^*$ (i.e., ASCII strings of \texttt{NULL}s and printable characters) endowed with edit distance;
	\item $G = \text{cyclic shifts on individual characters}$;
	\item Goal: remove \texttt{NULL}s and punctuation and make lowercase.
	\end{itemize}
\item Toy example 2:
	\begin{itemize}
	\item $X = \{97,\dots,122\}^*$ (i.e., lowercase alphabetical ASCII strings) endowed with edit distance;
	\item $G = \texttt{rot13}$ (or $G = 97 + 13 \cdot \mathbb{Z}/2\mathbb{Z}$);
	\item Goal: minimize the number of character edits plus the Hamming distance between the eventual result and the \texttt{rot13} of its reversal.
	\footnote{Note that strings such as $\texttt{gnat}$, $\texttt{tang}$, $\texttt{robe}$, $\texttt{serf}$, $\texttt{thug}$, etc. are left unaffected by this goal.}
	\end{itemize}
\item {\bf Language-theoretical security} \cite{momot2016seven}:
	\begin{itemize}
	\item $X = \text{files in a fixed format}$ endowed with a distance on, e.g. ASTs (see \S \ref{sec:AST});
	\item $G = \text{sugar-neutral bidirectional transformations}$;
	\item Goal: eliminate syntax that does not conform to reasonably specified deterministic grammar (possibly including syntax features such as \cite{lucks2017taming, ganty2017language}). 
	\end{itemize}
\item Feature elimination in C:
	\begin{itemize}
	\item $X = \text{C source files}$ with distance defined on ASTs;
	\item $G = \text{sugar-neutral translations}$ (between, e.g., source, LLVM, etc.);
	\item Goal: parsimoniously eliminate a particular type of syntactic sugar or other language feature.
	\end{itemize}
\item Binary patching:
	\begin{itemize}
	\item $X = \text{disassembled binaries}$ (for distance, see \S \ref{sec:FileTransformations});
	\item $G = \text{sugar-neutral lifts/translations/etc.}$;
	\item Goal: parsimoniously patch a known vulnerability.
	\end{itemize}
\end{itemize}

\subsubsection{\label{sec:DetailedExample}Detail of toy example 1}

Consider the alphabet $\mathcal{A} := \{0,32,\dots,126\}$ corresponding to ASCII \texttt{NULL} and printable characters, and let $X := \oplus_{j=1}^\infty A_{(j)}$ be the categorical direct sum of copies of $\mathcal{A}$, i.e., the infinite sequences over $\mathcal{A}$ with only finitely many nonzero entries. The string or word $w$ corresponding to an element of $X$ is defined here simply by removing any (trailing) zeros and mapping the numbers to ASCII characters. We endow $X$ with the Levenshtein distance $d_X$: i.e., edit distance with unit-cost insertions/deletions/substitutions. Since $|\mathcal{A}| = 96$, let $\mathcal{G} := \mathbb{Z}/96\mathbb{Z}$ act on $\mathcal{A}$ via cyclic unit shifts, and define $G := \prod_{j=1}^\infty \mathcal{G}_{(j)}$. Finally, define $P := X \times G$ to be the trivial ``principal $G$-bundle'' over $X$. 

The goal is to remove non-trailing \texttt{NULL}s and punctuation, and to make the input lowercase. This can be done by different combinations of atomic horizontal and vertical steps. Here, an atomic horizontal step means replacing $x \in X$ with $x' \in X$ with $d_X(x,x') = 1$, and an atomic vertical step means applying $g_k \in G$ of the form $g_k := \oplus_{j=1}^\infty \delta_{jk}$, where $\delta_{jk}$ equals the identity in $\mathcal{G}$ unless $j = k$, in which case it equals the unit shift in $\mathcal{G}$ that sends $0 \mapsto 32, 32 \mapsto 33, \dots, 125 \mapsto 126, 126 \mapsto 0$. 

It is easy to see that the optimal solution (in terms of number of steps) is to delete and insert characters. This is because \texttt{A} \dots \texttt{Z} and \texttt{a} \dots \texttt{z} respectively correspond to the \texttt{uint8}s $(65,\dots,90)$ and $(97,\dots,122)$, and the cost of making a character lowerspace by group actions is therefore $97 - 65 = 32 = 122 - 90$, whereas the cost is only 2 to lower case via a deletion followed by an insertion. 

But suppose we change the notion of an atomic vertical step to $g'_k := 31 \oplus_{j=1}^\infty \delta_{jk}$. Because 31 is coprime to 96, these atomic steps generate a group isomorphic to $G$, which provides a measure of justification for this change. We have that $31 \delta_{jk}$ sends $0 \mapsto 61, 32 \mapsto 62, \dots, 95 \mapsto 126, 96 \mapsto 0, 97 \mapsto 32, \dots, 126 \mapsto 60$. Now consider the initial string $\alpha = \texttt{ABCD}$: deletion and insertion to arrive at the goal $\omega = \texttt{abcd}$ requires a cost of 8, whereas applying $g'_1, \dots, g'_4$ yields $\beta = \texttt{`abc}$ at a cost of 4. Deleting the leading \texttt{`} and inserting a trailing \texttt{d} incurs an extra cost of 2, for a total cost of $6 < 8$. That is, the deletion/insertion strategy is not always optimal anymore, though it usually still will be. 

Observe that there is a ``flat connection'' on $P$, which is to say that the trivial factorization of $P = X \times G$ into horizontal and vertical spaces is $G$-equivariant.  That is, for all $x, y \in X$ there exists $g \in \pi^{-1}(x) \cong G$ such that $gx = y$.
\footnote{
For example, if $x = \texttt{ABCD}$ and $y = \texttt{ACD}$, then the corresponding $g = (0,1,1,-100)$ in ASCII numbering or $g = (0,1,1,-37)$ in ``internal'' numbering. 
}
In other words, the trivial factorization into horizontal and vertical spaces throughout $P$ should be taken as compatible with exchanging vertical displacements into horizontal ones:
\begin{equation}
\underbrace{d_G(x,gx)}_\text{vertical displacement} = \underbrace{d_X(x,y).}_\text{horizontal displacement} \nonumber
\end{equation}

To encode the idea of a collection of related files, we can use a \emph{local section} of $P$, which is a map $s: Y \rightarrow P$, where $Y \subseteq X$ is open (for the implicitly assumed discrete topology, this is a trivial restriction) and such that $s(y)_X := \pi(s(y)) = y$ for all $y \in Y$.  Moreover, goal-directed transformations are generally (lifts of) \emph{paths} $t: \mathbb{Z} \rightarrow Y$.  A path is is a \emph{parallel path} if $u(n)_G t(n) = t(n+1)$, for all $n \in \mathbb{Z}$, where here $u := s \circ t$ and the subscript $_G$ denotes the projection from the trivial bundle $P = X \times G$ onto its second factor. Notice that this allows the geometry of $X$ to be lifted into $G$ (or $P$) via the equation 
\begin{equation}
d_G(u(n)_G,id_G) := d_X(t(n),t(n+1)). \nonumber
\end{equation}

With the definition $L(t) := \sum_{n = -\infty}^\infty d_X(t(n),t(n+1))$ for \emph{path length}, we obtain the following

\

\textsc{Proposition.} If $t$ is a parallel path, then $L(t) = \sum_{n = -\infty}^\infty d_G(u(n)_G,id_G)$, i.e., the horizontal and vertical distances along a path are equal if the path is parallel. $\Box$

\

Armed with this, we can study problems of finding minimal-length paths subject to constraints on the path in $X$, e.g., specified initial and terminal points, or more saliently the requirement to stay within a specified subspace of $X$. As examples of the latter sort of constraint, consider a forbidden subsequence such as (the \texttt{uint8}) representation of \texttt{EOF} or a ``buffer'' constraint on the maximum number of nonzero entries of points in $X$. Such a restriction to some $Y \subseteq X$ entails the presence of more globally parallel sections and fewer locally parallel sections than on $X$, and this can be measured via relative sheaf cohomology, e.g. by comparing the cohomology to the space $\Gamma(X)$ of parallel sections on $X$.
\footnote{
In the event that $G$ is abelian, relative sheaf cohomology is comparatively straightforward: in the more general nonabelian case, one can either panic or contemplate much more abstract techniques such as outlined in \S 7 of \cite{lurie2009higher} (see also \cite{nlab:cohomology}).
}
\footnote{
Equivalences of files in $X$ induce additional highly nontrivial topological structure on the resulting quotient space, which can in turn induce curvature in \emph{any} connection (via Chern-Weil theory, for which see, e.g. \cite{milnor1974characteristic,rosenberg1997laplacian} and whose most basic incarnation is the Gauss-Bonnet theorem that relates the integral of the Gaussian curvature over a closed orientable surface to the Euler characteristic).
}

\subsubsection{\label{sec:FileTransformationImplementations}A framework for implementations}

Language representatives can be transformed by lifting/parsing them into semantics (i.e., CSTs/ASTs), transforming the semantics at a higher degree of abstraction, and projecting/unparsing them as in \cite{cousot2002systematic},
which points out that
\begin{quote}
[files] can be considered as an abstraction of their semantics. For example the syntax of [files] records the existence of [objects] and maybe their type but not [the trace of a parser or renderer], as defined by the semantics. \footnote{Here we have replaced the words ``program'' and ``variable'' with ``file'' and ``object,'' respectively.}
\end{quote}

In general, an AST will be subject to additional processing in order to reason over the syntax, frequently by producing an augmented AST and identifying vertices in some way to produce a suitably annotated digraph that we call a \emph{derived syntax graph} (DSG). 
\footnote{
Note that dependencies between (versus within) files of the same format suggest that we focus attention on properties relative to certain subspaces of the base space, with all the topological baggage that implies. It would be unusual except in very simple cases for the semantic base space to be topologically trivial. In the same vein, the strongest topological condition that seems likely to be broadly applicable to a bundle in the present context is that its universal cover is homotopy equivalent to a trivial bundle \cite{1993162}.
}
For example, the AST produced by a PDF parser will have nodes for indirect objects and their corresponding cross references (i.e., the byte offsets in the \texttt{xref} table) as well as some additional relevant information in the PDF trailer, and these data are subsequently associated to each other in a DSG, even if implicitly \cite{whitington2011pdf}. 

\begin{figure}
\centering
\begin{subfigure}[b]{.4\columnwidth}
\begin{small}
\begin{verbatim}
 1 START
 2 do while b
 3   do while b
 4     do while b
 5       do while b
 6         S
 7       enddo
 8       S
 9     enddo
10     if b
11       do while b
12         S
13       enddo
14       if b
15         S
16       endif
17     endif
18   enddo
19 enddo
20 HALT
\end{verbatim}
\end{small}
\caption{A toy program skeleton. Each \texttt{S} is its own statement or subroutine; each \texttt{b} is its own Boolean predicate.}
\end{subfigure} \quad 
\begin{subfigure}[b]{.5\columnwidth}
\includegraphics[trim = 60mm 100mm 45mm 100mm, clip, width=\columnwidth, keepaspectratio]{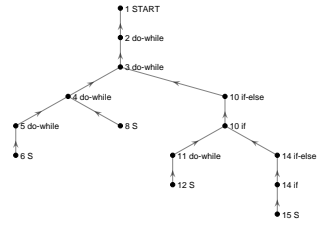}
\caption{The corresponding AST.}
\vspace{2ex}
\includegraphics[trim = 60mm 100mm 45mm 100mm, clip, width=\columnwidth, keepaspectratio]{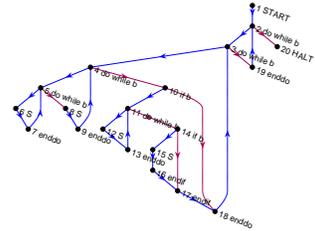}
\caption{The corresponding control flow graph. Branches are colored according to whether or not their corresponding \texttt{b} evals to ${\color{blue}\top}$ or ${\color{red}\bot}$.}
\end{subfigure}
\caption{\label{fig:ProgramAstCfg}
When suitably annotated with concrete details, semantically richer representations of a file are better suited for performing transformations on the semantics itself. The result can be unparsed and/or syntactically transformed as appropriate. Here, the control flow graph is a semantic augmentation of the AST that includes cross-references, i.e., a DSG.
}
\end{figure}

As a more generally familiar example, a compiler will use the AST of a computer program to produce a DSG in the guise of a control flow graph (Figure \ref{fig:ProgramAstCfg}). While parsing an input to a CST is invertible (if it is actually performed), parsing an input directly to an AST (or transforming a CST to an AST) is obviously very far from invertible. However, the transformation from an AST to a DSG is generally (or with only minor annotations, can be made) invertible.

Our present considerations suggest a general principle by which to separate concerns in file transformations that is algorithmically favorable: compositionally 
\footnote{
I.e., category-theoretically, which strongly constrains the form of a DSG.
}
manipulate an appropriate DSG, then unparse it into its corresponding AST, where geometrical considerations can be more naturally \emph{and efficiently} accounted for. For example, a DSG may be restructured, decomposed, and locally perturbed, and only the corresponding local ASTs need be compared for geometrical purposes (see \S \ref{sec:FibrationGeometry}) such as determining convergence to a desired state, e.g. the elimination of nondeterministic syntax elements \cite{momot2016seven}
or local similarity to some reference file. Such an approach inherits any compositional properties of the DSG and can be viewed through the lens of a category of lenses \cite{foster2007combinators, spivak2019generalized} (cf. \S \ref{sec:Lens}).

As a concrete example, a compiler might demand that a control flow graph be defined and decomposed as in \cite{johnson1994program, huntsman2019multiresolution},
restructured using the algorithm of \cite{zhang2004using} 
(cf. \cite{yakdan2015no, yu2005reverse}),
and then recursively manipulate (e.g., deoptimize) subroutines only if/as needed. Extending this example, the structured program theorem \cite{bohm1966flow, harel1980folk}
implies that we can choose a normal form for the structured control flow in which only \texttt{if-else} and \texttt{while} (or optionally also \texttt{case/switch}) constructs are used (see also Fig. \ref{fig:NoBranches}). 
\footnote{
These considerations apply equally well to decompilation: for example the immediately preceding control flow constructs can be recognized from the graph structure of the structured control flow graph alone \cite{sharir1980structural}, 
thereby separating more mechanical issues of control flow and modularity from undecidable problems such as variable name and type inference.
}

\section{\label{sec:Fibrations}Fibrations}

A \emph{fibration} \cite{ghrist2014elementary, hatcher2002algebraic} is a generalization of a fiber bundle that retains desirable homotopy properties, i.e., homotopy-equivalent fibers and the \emph{homotopy lifting property} which states that for any $f$ and $\tilde f_0$ that make the outer square of the following diagram commute, there is a map $\tilde f$ making the entire diagram commute:
\begin{center}
	\begin{tikzpicture}[->,>=stealth',shorten >=1pt]
		\node (v01) at (0,0) {$Y$};
		\node (v02) at (0,-2) {$Y \times [0,1]$};
		\node (v03) at (2,0) {$P$};
		\node (v04) at (2,-2) {$X$};
		\path [right hook ->] (v01) edge node [left] {$id \times \{0\}$} (v02);
		\path [->] (v01) edge node [above] {$\tilde f_0$} (v03);
		\path [->] (v02) edge node [below] {$f$} (v04);
		\path [->] (v03) edge node [right] {$\pi$} (v04);
		\path [style=dashed, ->] (v02) edge node [above] {$\tilde f$} (v03);
	\end{tikzpicture}
\end{center}

As a consequence, a path in $X$ can be uniquely lifted to a path in $P$. For example, in the context of homotopy type theory, dependent types are fibrations \cite{univalent2013homotopy}. A more general and abstract notion of fibration is provided the theory of model categories and homotopical algebra \cite{dwyer1995homotopy}.

\subsection{\label{sec:Lens}Lenses as bundles and fibrations}

Transforming files into a normal form has been considered as a mechanism to produce simpler, unambiguous (i.e., not polyglot \cite{raynal2010malicious}
or schizophrenic \cite{albertini2015funky})
files. However, complex structural dependencies such as checksums can obstruct \emph{ad hoc} solutions along these lines. The notion of a \emph{lens} \cite{foster2007combinators}
provides a principled, compositional solution that permits modifications to a file to be automatically transported to its putative normal form. Lenses have been synthesized at small scale from specifications and translation examples \cite{miltner2017synthesizing, maina2018synthesizing}, 
suggesting an approach for safely transforming files \cite{harris2019personal}.

It turns out that this lens-oriented approach can be fruitfully viewed from our perspective: indeed, a generalized lens category can be defined in terms of a category $\mathcal{C}$ and a functor $F : \mathcal{C}^\text{op} \rightarrow \mathbf{Cat}$ \cite{spivak2019generalized}. 
This recipe turns out to yield a \emph{Grothendieck fibration} or \emph{fibered category}, which can be thought of a generalized ``total space'' of a bundle (cf. \S \ref{sec:ModuliSpaces}). 
\footnote{
There are bundles (and similar objects) whose points are themselves bundles (and related objects), e.g., bundles of connections \cite{kobayaschi1957theory},
moduli spaces of bundles \cite{nlab:moduli_space_of_bundles},
etc. 
}
Indeed, many of the cases motivating the definition of this generalized lens category correspond specifically to bundles, and in particular bimorphic lenses can be interpreted as trivial bundles (i.e., the total space is a Cartesian product) \cite{spivak2019lenses}.

\subsection{\label{sec:ModuliSpaces}Moduli spaces}

As \cite{e2019machine} 
points out, 
\begin{quote}
A mathematically attractive definition of semantics is that it is the invariant after translation. If we view translation as operators between different [representations], the fact that semantics is preserved after translation means that the generators for different [representations] are all similar to one another [i.e., generators for representations commute with the corresponding translations].
\end{quote}
In other words, semantics is a \emph{modulus} (i.e., a complete isomorphism invariant) in the sense of algebraic geometry, wherein \emph{moduli spaces} or \emph{stacks} describe the algebraic invariants associated to \emph{categories fibered in groupoids} \cite{noohi}, 
\footnote{For the moduli stack of elliptic curves \cite{nlab:moduli_stack_of_elliptic_curves}, 
the appropriate (coarse, i.e., automorphism-forgetting) modulus is the $j$-invariant, which sends ``the'' modular curve $X(1)$ to the affine line; modular forms are sections of line bundles on this stack.
}
and wherein the role of ``total space'' is played by a Grothendieck fibration \cite{stacks-project}.

In the event that such ``generators'' and translations can be instantiated as linear operators, the spectra of the generators ought to be \emph{a priori} identical and yield ``semantic fingerprints.'' \cite{zhou2019mathematical} 
exploits this to perform high-performance unsupervised translation between natural language corpora. The essential step is to construct a Markov chain from statistics of the spacings between word pairs in a document, though other techniques (e.g., cross-correlations of tokens or words) might also be used in similar ways.

\subsection{\label{sec:FibrationGeometry}Geometry of program artifacts}

Transformations on dynamic program artifacts (e.g., ASTs, traces, error ontologies etc.) define relevant groupoids, and dissimilarity measures between these artifacts define relevant geometries on fibrations and their ilk. Here, we outline various (classes) of examples in this vein.

\subsubsection{\label{sec:AST}ASTs}

As suggested in \S \ref{sec:FileTransformations}, ASTs are well-suited for performing goal-directed transformations on files using a dissimiliarity measure (or outright metric) as an explicit objective function. 

For instance, edit distances for ASTs would be considerably less computationally expensive than edit distances for DSGs.
\footnote{
Tree edit distances are still quite expensive: the best known algorithm for edit distance on rooted labeled ordered trees requires cubic time \cite{demaine2009optimal}, 
and a subcubic algorithm is unlikely to exist \cite{bringmann2018tree}, 
though there are reasonably good practical algorithms \cite{pawlik2015efficient, pawlik2016tree} 
and approximations \cite{torsello2001efficiently, paassen2018tree}.
}
Also in this particular vein, tree edit distance is appealing due to the compositional structure of dynamic programs \cite{moor1994categories}
that compute it: i.e., edit distances are recursively computed from edit distances of substructures. Moreover, node annotations/labels can be taken into account in a way that separates their concerns from the tree structure by considering dissimilarities on \emph{attributed trees}. There are several potential avenues to producing a suitable and generic dissimilarity in this vein, e.g. combining known ordered tree isomorphism algorithms with the polytime approach of \cite{torsello2005polynomial}
for attributed rooted labeled unordered trees. Another avenue is to use kernels for attributed trees \cite{aiolli2015efficient, duessel2018detection}. 
\footnote{
It might be sufficient or even practically necessary to use the trivial metric on node annotations defined by $d(x, y) := 1$ for $x \ne y$, and $d(x, x) := 0$ for all $x$. Besides completing the specification of a dissimilarity measure, it seems likely that this particular case would admit tailored improvements relative to generic node annotation metrics.
}

\subsubsection{\label{sec:Traces}Traces}

An execution trace of a parser is a more dynamic and architecture-specific (i.e., operational \cite{nielson2007semantics}) 
representation of semantics than an AST. Considering traces as paths on the control flow graph of a program, one might coarse-grain subroutines \cite{huntsman2019multiresolution}
or roughly equivalently, use the dynamic sequence of function calls to get a suitably high-level notion of trace to define a relevant notion of algebra on individual fibers and a geometry relating fibers. A particularly useful class of dissimilarities on traces can be constructed using \cite{huntsman2015debruijn}.

In other words, each intermediate representation (token sequence, CST, AST, etc.) that a parser constructs defines a section in a fibration associated to a set of execution traces. 
Due to software errors, this section is typically local, but ideally global.

\subsubsection{\label{sec:Ontologies}Ontologies}

An order metric \cite{joslyn2010order}
can be applied across multiple instantiations of parser (or more generically, program) errors. This has the advantage that we can perform topological differential testing \cite{huntsman2020topological}
in concert with an error ontology to define a nice notion of \emph{de facto} syntax and get a reasonable notion of the ``base space'' $X$.

\subsubsection{\label{sec:Wasserstein}Generalized Wasserstein metrics}

A particularly interesting and general avenue for defining \emph{bona fide} metrics using category theory is suggested by \cite{patterson2019hausdorff},
which shows how to define a generalized Wasserstein metric on functors from a given small category to $\mathbf{Set}$. If for example we take the small category to be given by two parallel morphisms between two objects, such functors are quivers/multidigraphs. 

With a suitable functor, the Wasserstein metric applies to attributed quivers, and may admit specialization to be more narrowly tailored for a metric dissimilarity of a domain-specific form. One advantage of this approach is that the metric is a convex relaxation of a Hausdorff-style metric that admits computation via a linear program. Thus the algorithmic effort can be concentrated in the selection of an appropriate category and specialization of the linear program formulation to support fast evaluation.

\section{\label{sec:Remarks}Remarks}

As \S \ref{sec:DetailedExample} shows, even slightly nontrivial examples are intractable to explicitly analyze with this framework, but this is not its \emph{raison d'\^{e}tre}. Rather, this framework is intended to provide a conceptual basis for engineering transformation frameworks as in \S \ref{sec:FileTransformationImplementations}. Insisting on analogies or formal identifications with bundle-like objects endowed with geometry can inform the design and implementation of goal-directed file transformations.

%
%
%

%


\bibliographystyle{ieeetr}
\bibliography{GeoSSbib}

\end{document}